\documentstyle[graphics,latexsym,psfig]{mn}

\def\ros{{\sl ROSAT}}
\def\etal{{et\,al.}}

\def\msun{M$_{\odot}$}

\def\ergcm{\hbox{erg cm$^{-2}$ s$^{-1}$ }}
\def\ergs{\hbox{erg s$^{-1}$ }}
\def\degs{\ifmmode ^{\circ}\else$^{\circ}$\fi}
\def\amin{\ifmmode ^{\prime}\else$^{\prime}$\fi}
\def\asec{\ifmmode ^{\prime\prime}\else$^{\prime\prime}$\fi}
\def\fss{\hbox{$.\!\!^{\rm s}$}}        
\def\fdg{\hbox{$.\!\!^\circ$}}          
\def\farcs{\hbox{$.\!\!^{\prime\prime}$}}  
\def\h{$^{\rm h}$}\def\m{$^{\rm m}$}
\def\s{$^{\rm s}$}
\newbox\grsign \setbox\grsign=\hbox{$>$}
\newdimen\grdimen \grdimen=\ht\grsign
\newbox\laxbox \newbox\gaxbox
\setbox\gaxbox=\hbox{\raise.5ex\hbox{$>$}\llap
     {\lower.5ex\hbox{$\sim$}}}\ht1=\grdimen\dp1=0pt
\setbox\laxbox=\hbox{\raise.5ex\hbox{$<$}\llap
     {\lower.5ex\hbox{$\sim$}}}\ht2=\grdimen\dp2=0pt
\def\gax{\mathrel{\copy\gaxbox}}
\def\lax{\mathrel{\copy\laxbox}}

\def\sa18{SAX J1810.8--2609}
\def\r18{RX J1810.7--2609}

\begin{document}

  \title[SAX J1810.8--2609]{X-ray and optical\thanks{Partly based on 
    observations at the 
    European Southern Observatory, La Silla, Chile.}-to-infrared\thanks{The 
    United Kingdom Infrared Telescope is operated by the Joint Astronomy Centre
    on behalf of the U.K. Particle Physics and Astronomy Research Council.}
    follow-up   observations of the transient X-ray  burster \sa18}

  \author[Greiner et al.]{J. Greiner$^1$, A.J. Castro-Tirado$^{2,3}$,
          Th. Boller$^4$, H.W. Duerbeck$^5$,
          S. Covino$^6$, \\
          {\LARGE G.L. Israel$^7$,
         M.J.D. Linden-V{\o}rnle$^8$ and X. Otazu-Porter$^9$} \\
     $^1$ Astrophysical Institute
        Potsdam, An der Sternwarte 16, D-14482 Potsdam, Germany \\
     $^2$ Laboratorio de Astrof\'{\i}sica Espacial y F\'{\i}sica
              Fundamental (LAEFF-INTA),
              P.O. Box 50727, E-28080 Madrid, Spain \\
     $^3$ Instituto de Astrof\'{\i}sica de Andaluc\'{\i}a (IAA-CSIC),
        P.O. Box 03004, E-18080 Granada, Spain \\
     $^4$ Max-Planck-Institute for extraterrestrial Physics, Giessenbachstr. 1,
        D-85740 Garching, Germany \\
      $^5$ Free University Brussels, Pleinlaan 2, B-1050 Brussel, Belgium \\
      $^6$ Osservatorio Astronomico di Brera, Via Emilio Bianchi 46, 
        I-23807 Merate, Italy \\
      $^7$ Osservatorio Astronomico di Roma, Via dell'Osservatorio 2, 
        I-00040 Monteporzio, Roma, Italy \\
      $^8$ Niels Bohr Institute for Astronomy, Physics and Geophysics, 
        Astronomical Observatory, Juliane Maries Vej 30,
        DK-2100 Copenhagen, Denmark \\
      $^9$ Departament d'Astronomia i Meteorologia, Facultat de Fisica, 
        Universitat de Barcelona, Avgda. Diagonal 647, 08028 Barcelona, Spain
           }

   \date{Received 23 April 1999 / Accepted 1 July 1999 }

\pagerange{\pageref{firstpage}--\pageref{lastpage}}
\pubyear{1999}

 \maketitle

\label{firstpage}

\begin{abstract}
  We have performed a ROSAT follow-up observation of the X-ray transient
  \sa18\ on March 24, 1998 and detected a bright X-ray source (named
\r18) which was not detected during the ROSAT all-sky survey in September 1990.
Optical-to-infrared follow-up observations  of the 10\asec\ radius 
ROSAT HRI X-ray error box revealed one variable  object (R = 19.5
 $\pm$ 0.5 on 13 March, R $>$ 21.5 on 27 Aug 1998) which we tentatively
propose as the  optical/IR counterpart of \r18\ $\equiv$ \sa18.
\end{abstract}


\section{Introduction}

The $\sim$50 known X-ray bursters are thought to belong to the class of 
low-mass X-ray binaries (LMXB), i.e. their companions are thought to be 
low-mass ($\lax$0.5 \msun) late-type stars with $M_{\rm V} \gax 0$ (see e.g.
Lewin \etal\ 1995). The optical emission is usually dominated by
reprocessed X-ray radiation off the accretion disk and the contribution from
the secondary is generally negligible. Thus, if the X-ray emission in a LMXB
is transient, reprocessing results in a variable optical and IR source.

\sa18\ is a new transient in the Galactic bulge which was discovered 
with the Wide Field Camera 2 onboard BeppoSAX (Ubertini \etal\ 1998a).
The source has been detected on 1998 March 10 UT
at a level of 15 mCrab (2--9 keV), at R.A. = 18\h10\m46\s, Decl.
= --26\degr09\farcm1 (equinox 2000.0; preliminary error radius  of 3\amin).  
An X-ray burst of 45 sec duration was detected from this source on March
11.06634 with a peak intensity of about 1.6 Crab (Ubertini \etal\ 1998a). 
A BeppoSAX follow-up observation with the narrow-field instruments (NFI)
on 1998 March 12 detected the source at a flux level of 
1.5$\times$10$^{-10}$ \ergcm (corresponding to 7 mCrab) with a power law 
spectrum with
photon  index --2.2 (Ubertini \etal\ 1998b). The error radius of 1\amin\ 
of the originally published NFI position was later corrected to 3\amin\
(Ubertini \etal\ 1999) because the aspect solution for that NFI 
target-of-opportunity observation (TOO)  was particularly unfavourable 
due to the lack of bright stars in the pointing direction (a rather rare 
aspect control configuration).

Here we report follow-up observations of \sa18\ with the aim of
identifying its optical/infrared counterpart:
with ROSAT in X-rays to improve the position; immediate  optical/infrared 
observations to search for fading emission and later some deep imaging
to possibly identify the donor.

   \begin{figure}
    \vbox{\psfig{figure=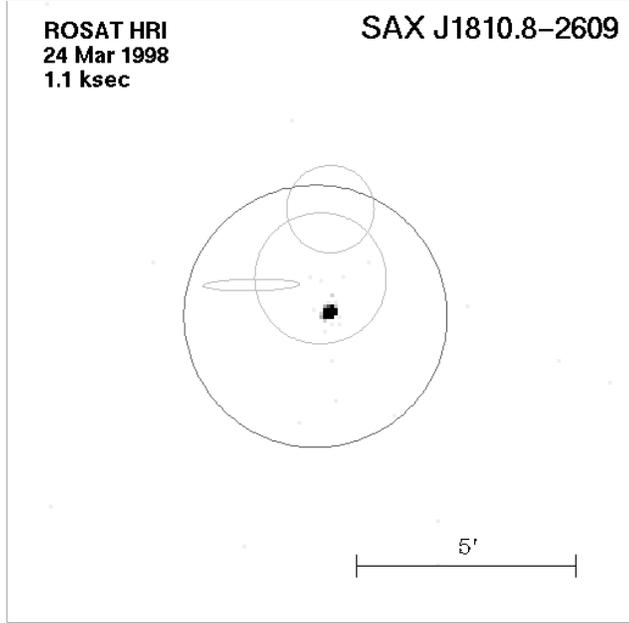,width=8.4cm,%
           bbllx=3.1cm,bblly=10.5cm,bburx=18.8cm,bbury=26.1cm,clip=}}\par
    \caption[rosfc]{The inner part of the HRI field of view with the strong
      X-ray source \sa18\ $\equiv$ \r18. The large, 3\amin\ radius circle 
      which is nearly centered around \r18\ is the BeppoSAX WFC error circle
      (Ubertini \etal\ 1998a), while the two smaller circles are the
      original (Ubertini \etal\ 1998b; top) and updated BeppoSAX NFI error
     circles (Ubertini \etal\ 1999; bottom). The ellipse marks the error
     box of IRAS 18077-2609.  North is up and East to the left. 
       No other X-ray sources are detected within 10\amin\ of \r18.
             }
      \label{rosfc}
   \end{figure}

\begin{table}
\caption{ROSAT observations of the \sa18\ location before and after the
  recent outburst}
\vspace{-0.1cm}
\begin{tabular}{crrrc}
\hline
\noalign{\smallskip}
   Date & Obs-ID$^{(1)}$ & T$_{\rm exp}$ & offaxis & CR$^{(2)}$  \\
   &  & (sec) & angle & (cts/s)   \\
\noalign{\smallskip}
\hline
\noalign{\smallskip}
 Sep. 12--16, 1990 & --~~~~~P & 230 & 0-55\amin & $<$0.029  \\      
 Sep. 10, 1993     & 400396P & 2\,014 & 52\amin & $<$0.027 \\
 Mar. 24, 1998     & 180267H & 1\,153   & 0\farcm25 & 0.54$\pm$0.02 \\
\noalign{\smallskip}
\hline
\end{tabular}

\noindent{\small

 $^{(1)}$  The letter gives the ROSAT detector 
           identification: P $\equiv$ PSPC, H $\equiv$ HRI. \\
 $^{(2)}$  Count rates are calculated for the 0.1--2.4 keV range (= PSPC
           channels 11-240). Upper limits are 3$\sigma$ confidence level.}
\label{xlog}

\bigskip
\caption{Optical-to-infrared observations of \sa18}
\vspace{-0.1cm}
\begin{tabular}{ccccc}
\hline
\noalign{\smallskip}
  $\!\!$ Date (1998) $\!\!$ & UT & Telescope & Filter & T$_{\rm exp}$ (sec) \\
\noalign{\smallskip}
\hline
\noalign{\smallskip}
 Mar. 13 & 07:26  & 1.5-m Danish &  Gunn-z  &         240          \\
 Mar. 13 & 07:30  & 1.5-m Danish &     R    &         240          \\   
 Mar. 13 & 07:39  & 1.5-m Danish &     B    &         240          \\  
 \,Jul. 25 & 03:01  & 1.5-m Danish &  Gunn-z  &         300          \\
 Aug. 27 & 00:05  & 1.5-m Danish &     B    &         240          \\
 Aug. 27 & 00:31  & 1.5-m Danish &  Gunn-i  &         240          \\
 Aug. 27 & 00:50  & 1.5-m Danish &     R    &         240          \\
 Aug. 28 & 23:22  & 1.5-m Danish &  Gunn-z  &         240          \\
 Sep. 15 & 23:05  & 0.9-m Dutch  & $\!\!$B/R/Gunn i$\!\!$ & 
   120/60/60$\!\!$      \\
\noalign{\smallskip}
\hline
\noalign{\smallskip}
 Aug. 14 & 09:15  & 3.8-m UKIRT  &     J    &         200          \\
 Aug. 14 & 09:30  & 3.8-m UKIRT  &     K    &         100          \\
\noalign{\smallskip}
\hline
\end{tabular}
\label{olog}
\end{table}

\section{ROSAT Observations}

A target-of-opportunity observation with the ROSAT HRI was performed on
1998 March 24 of the error box of \sa18. The total exposure time was 1153 sec
and obtained within one observation interval (22:14:46--22:33:59 UT).
One strong X-ray source was detected at a mean count rate of 
0.54$\pm$0.02 cts/s. The best-fit X-ray
position is R.A. = 18\h10\m44\fss5, Decl. = --26\degr 09\amin01\asec\
(equinox 2000.0), with an error radius of  $\pm$10\asec\ (comprising
a 5$\sigma$ statistical error of 2\farcs5 and a systematic bore sight 
uncertainty of 8--9\asec). 
This ROSAT position is not consistent with the only cataloged source
IRAS 18077-2609 as mentioned in Ubertini \etal\ (1998a; see also 
Fig. \ref{rosfc}).

In view of the original discrepancy between this ROSAT HRI position and
the BeppoSAX NFI position the following checks have been performed on the
ROSAT attitude data (courtesy W. Grimm/GSOC):
\begin{itemize}
\vspace{-0.25cm}\item the log-file of the attitude computation does not 
  reveal any irregularities
\item the locations of the minimum number of three optical 
  stars was always available
\item the distribution of the residua between expected and 
  measured attitude for each time bin is identical to the mean value of the
  last 7 years
\item during the observation the attitude wobbles
  by $\pm$2\amin\ (as expected) around the mean value of R.A. = 272\fdg2947 
  and     Decl. = --26\fdg1503 (equinox 2000.0)
\item the catalog positions of the three optical stars
  used in the attitude determination have been checked against the PPM catalog 
  and were found to be accurate to $\pm$1\asec.
\end{itemize}\vspace{-0.15cm}
The main conclusion is that the attitude during this ROSAT observation of 
\sa18\ is in no way abnormal as compared to those derived over the last
7 years. The subsequent solution of the ROSAT-BeppoSAX discrepancy by
recognizing a rare aspect control configuration of the BeppoSAX satellite 
during their TOO leads to the conclusion that the bursting and persistent 
source seen by the BeppoSAX satellite and the source \r18\ seen by
ROSAT are the same source.

The X-ray intensity of \r18\ during the ROSAT HRI observation does not vary
by more than a factor of 3, and no X-ray burst is detected during the
observation.
No coherent or quasi-periodic oscillations are found in the 2--200 sec range,
yielding a 3$\sigma$ upper limit of the pulsed fraction of 40\%.
Overall, the X-ray light curve is consistent with a constant source.

   \begin{figure*}
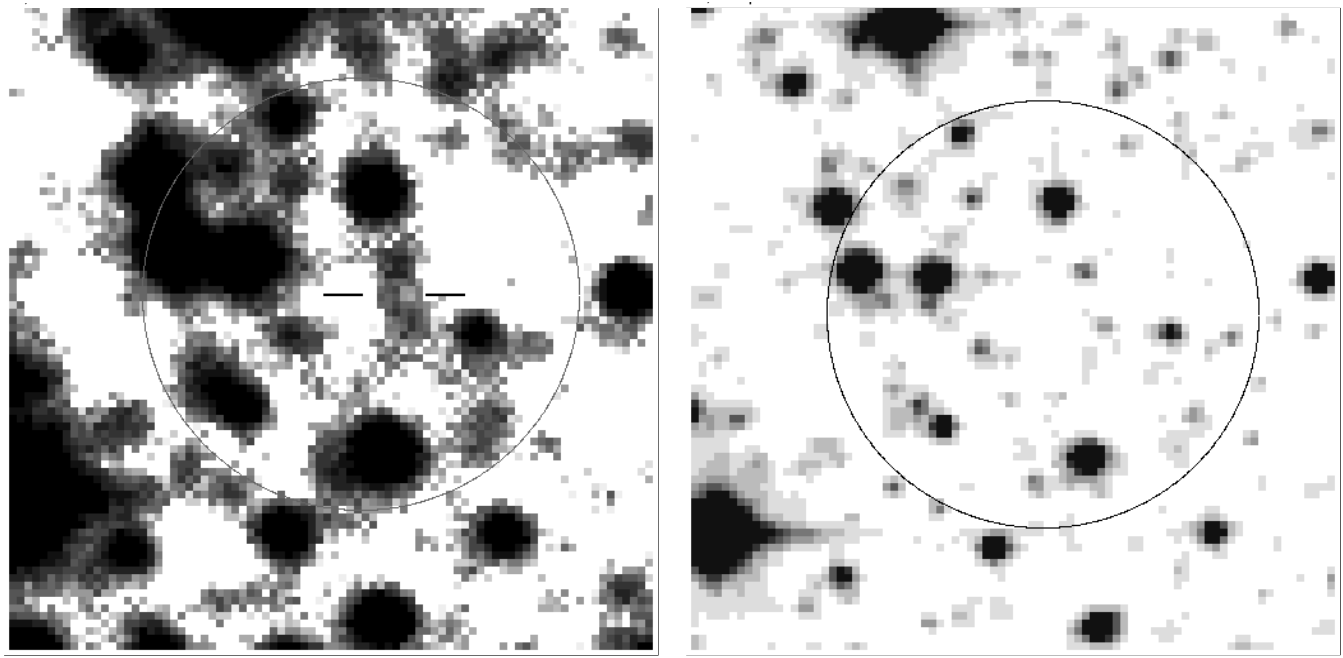

      \vbox{\psfig{figure=sax1810_saxzf.ps,width=8.8cm,%
           bbllx=3.1cm,bblly=10.5cm,bburx=18.8cm,bbury=26.2cm,clip=}}\par
      \vspace*{-8.75cm}\hspace*{9.cm}
      \vbox{\psfig{figure=sax1810_saxz3.ps,width=8.8cm,%
           bbllx=3.1cm,bblly=10.5cm,bburx=18.8cm,bbury=26.2cm,clip=}}\par
      \caption[fc]{A 40\asec $\times$ 40\asec\ field containing the 
          10\asec\ radius X-ray error box overlayed on the 
          Gunn-z band images obtained on 13 March (left) 
          and 28 August 1998 (right). North is up and East to the left. 
          The variable object and tentative counterpart to \sa18\ is
          marked with two dashes in the left image.
             }
      \label{fc}
   \end{figure*}

The location of \sa18\ was covered by the ROSAT all-sky survey in September
1990 for a total of 230 sec. No source was detected at that time, yielding
a 3$\sigma$ upper limit of 0.029 PSPC cts/s. In addition, the \sa18\ 
location is also in the field of view (though very far off-axis) of the 
pointing 400396 (PI: C. Motch) on 10 September 1993. The roughly 2 ksec 
exposure results in a 3$\sigma$ upper limit of 0.027 PSPC cts/s.
A summary of the relevant numbers is given in Tab. \ref{xlog}.
These non-detections at a sensitivity level of a factor of 20 below
the count rate detected during the HRI TOO (note that the HRI is a factor
of about 3 less sensitive for hard-spectrum X-ray sources than the PSPC) 
proves the transient nature of the detected ROSAT HRI source \r18.

Despite its location near the Galactic plane (b$^{\rm II}$=--3\fdg5), the 
source \sa18\ is located in a direction of relatively low absorbing column: 
the total Galactic column is 
$N_{\rm H}^{\rm gal}$=3.7$\times$10$^{21}$ cm$^{-2}$ 
(Dickey \& Lockman 1990) corresponding to $A_{\rm V}^{\rm gal} = 2.7$ mag
(using the relation $A_{\rm V} = 17/23 \times N_{\rm H}$ [10$^{21}$ cm$^{-2}$]
from Predehl \& Schmitt 1995).
Using this absorbing column (together with the Morrison \& McCammon 1983
cross section model) and a power law model with
photon index --2.2 as derived from the BeppoSAX NFI pointing 
(Ubertini \etal\ 1998b) we find an unabsorbed flux of 
1.1$\times$10$^{-10}$ \ergcm  in the ROSAT band (0.1--2.4 keV) or
correspondingly 3.5$\times$10$^{-11}$ \ergcm in the 2--10 keV band -- i.e.,
a factor of 4 lower than during the BeppoSAX observations on Mar. 10-12, 1998
(Ubertini \etal\ 1998b). The luminosity during the HRI observation is
1.2$\times$10$^{36}$ (D/10 kpc)$^2$ \ergs (0.1--2.4 keV).

\section{Optical-to-infrared Observations}

Immediately after notification of the \sa18\ transient
images in the Bessel B and R and the Gunn z bands were taken on 1998 Mar. 13 
with the Danish 1.5-m telescope (+ DFOSC) at ESO's La
Silla (Chile) Observatory. 
Further imaging was obtained with the same instrument on July 25 and 
Aug 27--28, 1998. The full observing log is displayed in Table \ref{olog}.
The raw frames were processed with standard techniques within the 
IRAF\footnote{IRAF is
  distributed by National Optical Astronomy Observatories, which is
  operated by the Association of Universities for Research in Astronomy,
  Inc., under contract with the National Science Foundation.} 
environment, which included bias subtraction and flat-fielding (using flat 
field images obtained during twilight). The resultant B, R and Gunn i 
band frames were calibrated using T Phe and other stars in the standard field 
SA 110 (Landolt 1992). SEXtractor (Bertin \& Arnouts 1996) was used to
obtain the magnitude estimates.

\begin{table*}
\vspace{-0.15cm}
\caption{BRiJK magnitudes for local standards in the \sa18\ field}
\vspace{-0.20cm}
\begin{tabular}{cccccc}
\hline
\noalign{\smallskip}
Star ID &       B        &       R        &       i        &    J    &    K  \\
\noalign{\smallskip}
\hline
\noalign{\smallskip}
 A      & 19.06$\pm$0.10 & 16.04$\pm$0.06 & 14.66$\pm$0.06 & 12.93$\pm$0.04 &
    11.81$\pm$0.04      \\
 B      & 19.64$\pm$0.10 & 16.73$\pm$0.10 & 15.37$\pm$0.09 & 13.50$\pm$0.05 &
    12.56$\pm$0.05      \\
 C      & 19.46$\pm$0.10 & 16.99$\pm$0.08 & 15.76$\pm$0.10 & 14.17$\pm$0.06 &
    13.42$\pm$0.07      \\
 D      & 21.10$\pm$0.20 & 18.10$\pm$0.15 & 16.85$\pm$0.15 & 15.30$\pm$0.15 &
             --          \\
 E      &      --       & 18.50$\pm$0.20 & 17.50$\pm$0.20 & 15.30$\pm$0.15 &
    14.30$\pm$0.10        \\
\noalign{\smallskip}
\hline
\end{tabular}
\label{comp}
\end{table*}

Several objects are seen within the HRI 
error circle. But a comparison of the March 13 images with those obtained 
in July/August reveals only one object 
which is not seen in the deeper images obtained on 
Aug 27--28, 1998 (Fig. \ref{fc}; limiting magnitudes are
B $>$ 22.5 mag, R $>$ 21.5 mag, i $>$ 20 mag).
Though a direct numerical comparison of the frames taken in March and August,
respectively, is hardly possible due to different seeing conditions
(1\farcs6 on March 13 and 0\farcs9 on Aug. 27) and the non-circular
shape of the stellar images, this object is (1) the only new object
and (2) seen in three colors. For this new object 
we measure B = 21.5 $\pm$ 0.5 mag and R = 19.5 $\pm$ 0.5 mag, 
and a signal-to-noise ratio of 5 in the R and Gunn z band images
(we have unfortunately no flux calibration for the Gunn z band image).
Its position is R.A. = 18\h10\m44\fss4, Decl. = --26\degr 09\amin00\asec 
(equinox 2000.0), with an error radius of $\pm$1\asec.

\begin{figure}
   \resizebox{\hsize}{!}{\includegraphics{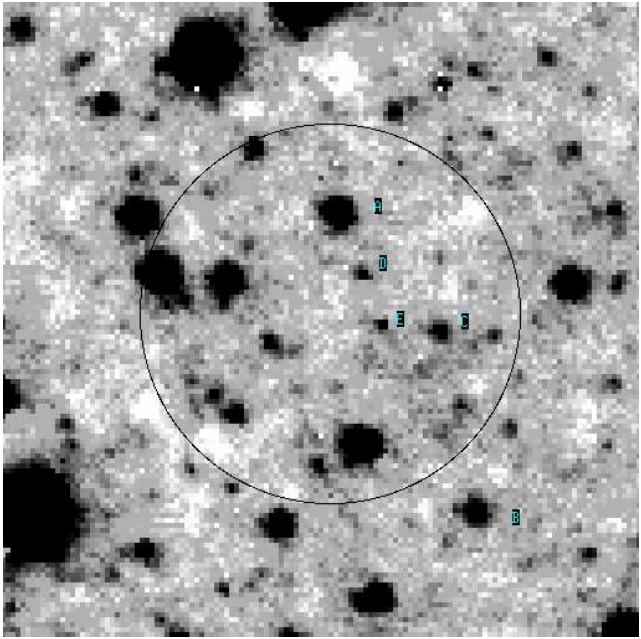}}
   \caption[kimab]{A 38\asec $\times$38\asec\ field in the K-band, as obtained 
                 at UKIRT on 14 Aug 1998. It includes the 10\asec\ radius 
                 ROSAT X-ray error box for \sa18. North is up and East 
                 to the left. Local standards are marked 
                 (Tab.\,\ref{comp}). 
             }
      \label{kima}
   \end{figure}

We also obtained service observations at the United Kingdom 3.8-m Infrared
Telescope (UKIRT) in Hawaii on August 14, 1998 after the more accurate ROSAT 
position had been determined. J and K-band frames were acquired using the
IRCAM3 infrared camera. IRCAM3 is a near-IR detector (1-5 $\mu$m) with
a 256 $\times$ 256 pixel InSb array. The image scale is
$\sim$0\farcs3/pixel giving a field of view of 73\asec$\times$73\asec. 
A 17$^{\prime\prime}$ dithering around the ROSAT coordinates was performed, 
thus obtaining five frames (40 s each in the J-band and 20 s each in the 
K-band) in order to determine and subtract the
background from each of the individual images. The resultant frames were 
calibrated using the standard star FS27 (Cassali \& Hawarden 1992). 
Fig. \ref{kima} shows a blow-up of one of these K-band images together with
the identification of various objects (see Tab.\,\ref{comp}). 
The object described above is not seen in these images, and
limiting magnitudes  are K $>$ 16.5 mag and J $>$ 17.5 mag (Aug 14). 

\section{Discussion}

Based on (1) the expectation of correlated X-ray/optical variability in
LMXBs as described in the introduction, and (2) our finding of only 
one variable object in
the optical/IR bands which moreover has a brightness change in the
direction as expected for an LMXB counterpart, we tentatively propose
this variable object as the
counterpart of the transient X-ray burster \sa18\ $\equiv$ \r18.

The number density of variable stars at 20th mag is poorly known. Simple 
scaling from extensive searches for variable stars in the 14--17 magnitude 
range suggests about 80 $\Box$$^{-2}$ brighter than 20th mag at low galactic
latitudes (e.g. Hudec 1998, Hudec \& Wenzel 1996). This implies a chance
probability of 2$\times$10$^{-3}$ to find an unrelated variable star 
inside the 10\asec\ ROSAT HRI error box, thus strengthening our conclusion
to have found the counterpart of \r18.

If we correct our brightness measurements for the total Galactic extinction of 
$A_{\rm V}^{\rm gal} = 2.7$ mag,
we obtain a color of B--R\,=\,0.4 mag during the March 1998 observation.
This is consistent with the expected color of a disk which is heated
by strong reprocessing and thus would have colors similar to a hot star: 
B--R $\approx -0.1$ mag.

The non-detection of persistent K-band emission during August 1998 is also
consistent with the expectation for the donor (assuming a quiescent system
in which the accretion disk does not dominate the light). With a ratio
$A_{\rm K}/A_{\rm V}$=\,0.112 (Rieke \& Lebofsky 1985) the total Galactic 
K-band extinction is $A_{\rm K}$=\,0.31 mag. With a location only 6 degrees
away from the Galactic Center, the distance of \sa18\ is likely of the order
of 8 kpc. Indeed, Cocchi \etal\ (1999) recently estimated
a distance of 5 kpc from one particularly bright X-ray burst which showed
signatures of photospheric radius expansion.
With this distance and using a mean (V--K)$_0$=1 mag (Johnson 1966), 
our K-band limiting magnitude of ${\rm K} > 16.5$ mag implies 
$M_{\rm V} > 4.1$ mag, consistent with a late-type main-sequence donor,
and similar to other X-ray bursters. With a range of $6 < M_{\rm V} < 10$ mag
for late-type main-sequence donors and intrinsic colours of
$2 < (V-K)_0 < 4.5$ mag the expected K band magnitude of the quiescent
donor is K $\sim$ 17.5--19 mag, thus making even spectroscopy feasible.

The amplitude of $\Delta$R $\gax$ 2 mag between March and August 1998
requires a difference in the X-ray illumination (flux) of about a factor of
100 (according to the relation between X-ray flux and reprocessed
visual flux found by van Paradijs \& McClintock 1994). Such a
factor in flux difference is reasonable to be valid for this transient 
(though we are not aware of an X-ray observation during August 1998).

We therefore conclude that the brightness and colors of the proposed
counterpart to \sa18\ obtained during two epochs in 1998 are consistent with 
the expectations for a transient X-ray burst source. Deeper follow-up infrared
photometry and further spectroscopy are necessary to validate 
our tentative identification.

\section*{Acknowledgements}

We thank J. Tr\"umper for granting TOO time with ROSAT which made this study
only possible. We are indebted to W. Grimm (German Space Operation Center, 
Oberpfaffenhofen) for a thorough analysis of the ROSAT attitude solution
as well as to P. Ubertini for extensive discussions
to clarify the original discrepancy with the BeppoSAX NFI position.
We also are grateful to S. Legget, J. Davies and the UKIRT staff for the
service observations.
JG is supported by the German Bundesmi\-ni\-sterium f\"ur Bildung,
Wissenschaft, Forschung und Technologie
(BMBF/DLR) under contract No. FKZ 50 QQ 9602 3.
The \ros\, project is supported by BMBF/DLR and the Max-Planck-Society.
This research has made use of the Simbad database, operated at CDS, 
Strasbourg, France.

\label{lastpage}
\end{document}